\documentclass[letterpaper,12pt]{report}

\title
{
	Design and Implementation of a Secure Web-Based File Exchange Server\\
	Specification Design Document\\
}

{\author
	{\bf
		{CIISE Security Investigation Initiative}\\\hline\\
		Represented by:\\\\
		Serguei A. Mokhov\\
		Marc-Andr\'e Laverdi\`ere\\
		Ali Benssam\\
		Djamel Benredjem\\
		\texttt{\{mokhov,ma\_laver,al\_ben,d\_benred\}@ciise.concordia.ca}
		\\\\\\
		Montr\'eal, Qu\'ebec, Canada\\\\\\
	}
}

\date{December 14, 2005}

\usepackage{graphicx}
\usepackage{latexsym}
\usepackage{makeidx}
\usepackage{url}

\makeindex

\newcommand{\lucidL}[1]{{$\mathit{Lucid}$}($L$) }

		{}

\def\myvert{\raise 2.27pt \hbox{\vrule depth 0pt height 8pt width 0.2mm}}
\def\myarrow{\hspace*{0.43mm}%
             \raise 2.29pt\hbox{\vrule depth 0pt height 8pt width 0.16mm}%
             \hspace*{-0.32mm}%
             $\longrightarrow$
             \ %
             }

\begin{document}

	\begin{titlepage}
		\maketitle
	\end{titlepage}

	\pagenumbering{roman}
\tableofcontents
\clearpage
\pagenumbering{arabic}


	\chapter{Introduction}
\index{Introduction}



Building Trust is the basis of all communication, especially
electronic one, as the identity of the other entity remains
concealed. To address problems of trust, authentication and security
over the network, electronic communications and transactions need a
framework that provides security policies, encryption mechanisms and
procedures to generate manage and store keys and certificates.

The Public Key Infrastructure (PKI) is a security architecture that
has been introduced to provide an increased level of confidence for
exchanging information over increasingly insecure networks, such as
the Internet. A PKI infrastructure is expected to offer its users a
secure and trustworthy electronic transaction.

\section {Purpose}
The intent of implementation and deployment of  PKI facilities is to
meet its basic purpose of providing Trust. Presently, PKI needs to
perform the following security functions:

\begin{itemize}

\item  \emph{Mutual authentication of entities taking part in the
communication:} Only authenticated principals can access files to
which they have privileges.

\item \emph{Ensure data integrity:} By issuing digital certificates
which guarantee the integrity of the public key. Only the public key
for a certificate that has been authenticated by a certifying
authority should work with the private key possessed by an entity.
This eliminates impersonation and key modification.

\item \emph{Enforce security:} Communications are more secure by using SSL to transmit information.

\end{itemize}

\section {Scope}
PKI is implemented to secure sensitive resources of the organization
and avoid security breaches. The PKI environment allows trustworthy
communication between the different principals. These principals
must be authenticated and the access to the resources (files) should
be secured and regulated. Any principal wants to access to the
database needs to perform the following steps:

\begin{itemize}

\item  \emph{Mutual authentication:} The Web Server via which the database is contacted authenticates the
principal using its digital certificate and username to ensure that
it is who it claims to be . The principal authenticates also the
server using its certificate information.

\item  \emph{Principal validation:} To validate the principal, the
server looks up information from an LDAP server which contains the
hierarchy of all principals along with certificates and credentials.
The LDAP service is compliant with the X.500 database structure.

\item  \emph{Enforcing security:} The security is enforced by using SSL to communicate between the
Web Server and the LDAP server, the Web Server and the database and
between the principal and Web Server.

\item  \emph{Principal authentication:} Upon successful authentication,
the Web Server will allow the principal to perform actions on the
database according to a pre specified Access Control List.

\item \emph{Kinds of users:} We distinguish between a normal and
an administrator. While a normal user can upload, download, delete
and view files; the administrator has the ability to: upload,
download, delete and view files; add, delete and modify users;
generate user's certificate, with all required information; generate
ACL to users; manage groups, perform maintenance.

\end{itemize}

Finally, this infrastructure allows additional features such as the
ability to assign users to groups in order to provide users with the
access to files prepared by other group members.

\section {Definitions and Acronyms}

\begin{itemize}
\item     PKI: Public Key Infrastructure
\item     OpenLDAP : is a free, open source implementation of the Lightweight
          Directory Access Protocol (LDAP).
\item     OpenSSL: an open source SSL library and certificate authority
\item     Apache Tomcat: A Java based Web Application container that was created to
          run Servlets and JavaServer Pages (JSP) in Web applications
\item     PostgreSQL: An open source object-relational database server
\item     SSL: Secure Socket Layer
\item     JSP: Java Server Pages
\item     JCE: Java Cryptography Extension
\item     API: Application Programming Interface
\item     JDBC: Java Database Connectivity
\item     JNDI: Java Naming and Directory Interface
\item     LDAP: Lightweight Directory Access Protocol
\item     X.509: A standard for defining a Digital Certificate used by SSL
\item     SRS: Specification request Document
\item     SDD: Specification Design document
\item     DER: Distinguished Encoding Rules
\item     Mutual Authentication: The process of two principals proving their identities to each other
\item     SFS: Secure File Exchange Server, this product
\item     COTS: Commercial Off The Shelf, common commercially or freely available software

\end{itemize}
 

%
%

\chapter{System Architecture}

This chapter is intended to provide an overview of the whole system
as proposed in the previous requirements and specification document.
It describes the product's perspective, interfaces and design
constraints as we have assumed.
We will first describe the architectural guidelines for this
product, followed by software interface design design, and hardware
environment.

\section{Architectural Philosophy}

The SFS technology hereby implemented is running on architecture
that provides a high level of secrecy and integrity for exchanging
information. The system is externally visible only through a web
application for normal users, and is also entirely visible and
accessible for administrators in the scope of normal operations. In
addition, the system assumes an internal certificate authority which
is explicitly trusted by all principals using SFS.

For the proposed architecture, it requires mutual authentication
between the user and the web server, an LDAP validation of the user
by using digital certificates, the use of the SSL protocol to
enforce the security over the communication between modules and the
preservation of files in a database.

This architecture must respect the following properties:
\begin{itemize}

\item \emph{Security:} The confidentiality, integrity and
availability of information. This is to be implemented by supporting
the data encryption and certificates mechanisms for secured
communication, as well as specifying an access control mechanism for
the files stored

\item \emph{Trustworthiness:} The use of electronic certificates internally generated,
and of specific use for the application, allow an high trust to be
given to the user.

\item \emph{Scalability:} SFS can be expanded easily to cope with large loads.
Methods such as load balancing and replication can be easily
integrated.

\item \emph{Openness:} The proposed architecture can be implemented and
deployed using Java Technologies and open source tools that are
well-used and rely on standards. The SFS itself is an open source
product developed to achieve security objectives.

\item \emph{Component-Based Software Engineering:}  The SFS
framework may be treated as components (modules) . We have already
mentioned the relevant technologies that can better fit for each of
these modules.

\item \emph{Usability:} The SFS service must be designed for high usability.
All the required information for a single operation should be
grouped in a single screen, with a minimum number of screens needed
for all the application.

\end{itemize}

This architecture aims at maximizing software reuse by the
integration of COTS applications, portability and interoperability
by the use of standards, security and scalability by the user of a
single access point.

\section{Components}

\begin{figure}[htbp]
\begin{center}
  \fbox{
      \scalebox{0.6}{  \includegraphics{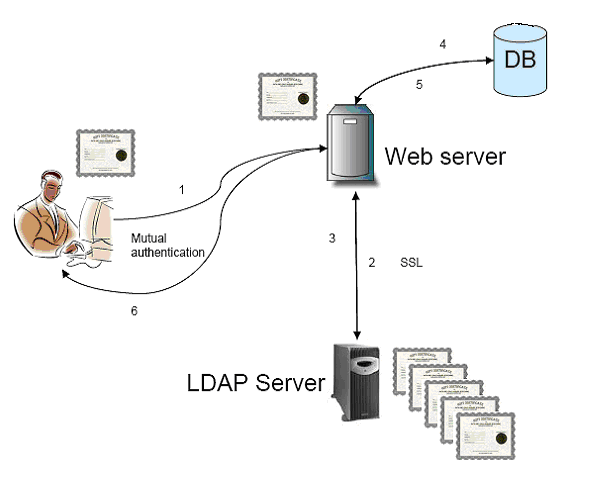}
      }
    }
    \caption{\label{fig:architecture} Main system architecture}
\end{center}
\end{figure}

The Figure \ref{fig:architecture} describes the overall architecture
of the SFS system. We see four major interacting components: the
user interface, the web server, the database and the LDAP server.
Two components are not displayed on this figure , which are the
certificate authority and the logging engine.

\subsection{User Interface} The user interface constitutes of HTML
web pages that the user uses through a web browser. Those web pages
are generated by the web server and interact exclusively with the
SFS web server.

\subsection{Web Server} The web server is the single access point
of the system. It handles authentication responsibilities, database
access and user interaction. This is to be handled by the Apache
Tomcat 5.0 server and custom J2SE 5.0 code.

\subsection{Database Server} This database server contains all
the information about the files and their access control rights. It
contains also a subset of the user information. This is to be
handled by PostgreSQL.

\subsection{LDAP Server} This specialized server holds the user
credentials (notably user name and password). It could be extended
to include user certificates. This module will be realized by
OpenLDAP.

\subsection{Certificate Authority} This responsibility is manually managed
by administrators. Using software tools, they are able to generate
the user and server certificates. In our case, we use OpenSSL to
perform those functions.

\subsection{Logging Engine} This component is responsible for collecting the audit
trails and debug information from other components and store it
locally. We wanted to use log4j, but we finally opted for the
logging mechanisms available in the tools we are using, notably by
using Tomcat's logging.

\section{Interactions}

We will now describe the inter-module interactions by the use of a
system scenario.

The user, with a web browser, connects to the web server using SSL.
The web server, being configured as to require client
authentication, both parties exchange their certificates and
validate their peer's identity. The web server then prompts the user
for a user name and password, thereby enforcing 2-factor
authentication.

Upon receiving this information, the web server queries the LDAP
directory based on the user name and retrieves the user's password
hash and certificate (if any is defined). The web server then
proceeds to hash the plaintext password (using SHA1) received from the user and compare with the one
from the LDAP server. If that information (as well as the
certificates, if any) matches, the user is logged in the system.

The web server will then query the database server for the access
rights of the user (administrator or normal user) and the list of
files the user has access to. Based on this information, it will
display the appropriate user interface functions and the file list.

On user requests to upload, delete or download files, the web server
will request the database server to perform the needed transactions.

\section{Software Interface Design}

This section describes the software interfaces (commonly referred to
as APIs) to be used to communicate between each component.

\subsection{Web Server} The web server is reached by the client
through an HTTPS connection to the single point of access for the
web application, the login screen.

\subsection{Database Server} The database server is reached
using the JDBC API with the official PostgreSQL JDBC driver.

\subsection{LDAP Server} The LDAP server is reached using
the default Sun LDAP JNDI driver. The LDAP protocol is used for the
queries and is encapsulated by JNDI.

\subsection{Certificate Authority} This component is not integrated
with others and, as such, does not have a software interface to
document.

\subsection{Logging Engine} On the Web server, this component is called
automatically by the use of the default output and default error
streams, which will write the information to a log instead of a
console.

\section{Hardware Environment}

The SFS system is expected to work in a networked environment,
possibly a LAN, but not necessarily. We assume that the principals
have a network connection allowing to communicate with each other.
Only typical low-end workstation hardware (such as a Pentium III system with
256+ MB of RAM, 2GB hard drive with a 10BaseT Ethernet connection)
is required to operate all the components of the system, which may
be distributed or centralized as needed. Ordinary P III with 256+ MB
of Ram and 2GB HDD are the minimum requirements needed to deploy the
system.

\section{Code View}

\begin{figure}[htbp]
\begin{center}
  \fbox{
      \scalebox{0.6}{  \includegraphics{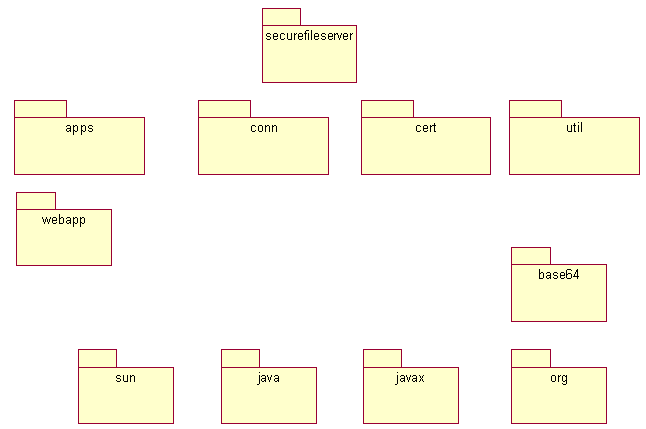}
      }
    }
    \caption{\label{fig:packages} Java Packages}
\end{center}
\end{figure}

We decided to structure our software in a few main packages,
 as illustrated in Figure \ref{fig:packages}.

Those packages hold as follows:

\begin{enumerate}
    \item \emph{securefileserver:} root of our custom code
    \item \emph{base64:} Library for encoding and decoding Base64-encoded data
    \item \emph{apps:}  Various applications
    \item \emph{webapps:}  The web application code
    \item \emph{conn:}  Connection abstraction code, such as SSL connections,
                                            LDAP connections and database connections
    \item \emph{cert:}  Certificate authority code
    \item \emph{util:}  Utility classes, such as the configuration file loader
\end{enumerate}

\begin{figure}[htbp]
\begin{center}
  \fbox{
      \scalebox{0.6}{  \includegraphics{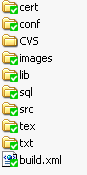}
      }
    }
    \caption{\label{fig:dir} Repository Structure}
\end{center}
\end{figure}

Since we had to deal with a large set of non-code artifacts, we
structured our repository as illustrated in Figure \ref{fig:dir}.s

Those directories hold as follows:
\begin{enumerate}
    \item root directory: contains the Ant makefile and all the other subdirectories
    \item \emph{cert:} certificates generated manually using OpenSSL
    \item \emph{conf:} configuration files
    \item \emph{CVS:}  repository management code, handled automatically by CVS
    \item \emph{images:} documentation-related images
    \item \emph{lib:}  libraries in JAR format
    \item \emph{sql:}  database initialization script
    \item \emph{src:} Java source code
    \item \emph{tex:}  documentation in \LaTeX2e format
    \item \emph{txt:}  various notes in text format
    \item \emph{design:}  Rational Rose model of the system
\end{enumerate}

\chapter{Detailed System Design}

In this section of the specification document we elaborate the
detailed description of the main modules and subprograms of the SFS
system. We provide the important class diagrams
for the different packages of the
design phase as mentioned above.

\begin{figure}[htbp]
\begin{center}
  \fbox{
      \scalebox{0.41}{\includegraphics{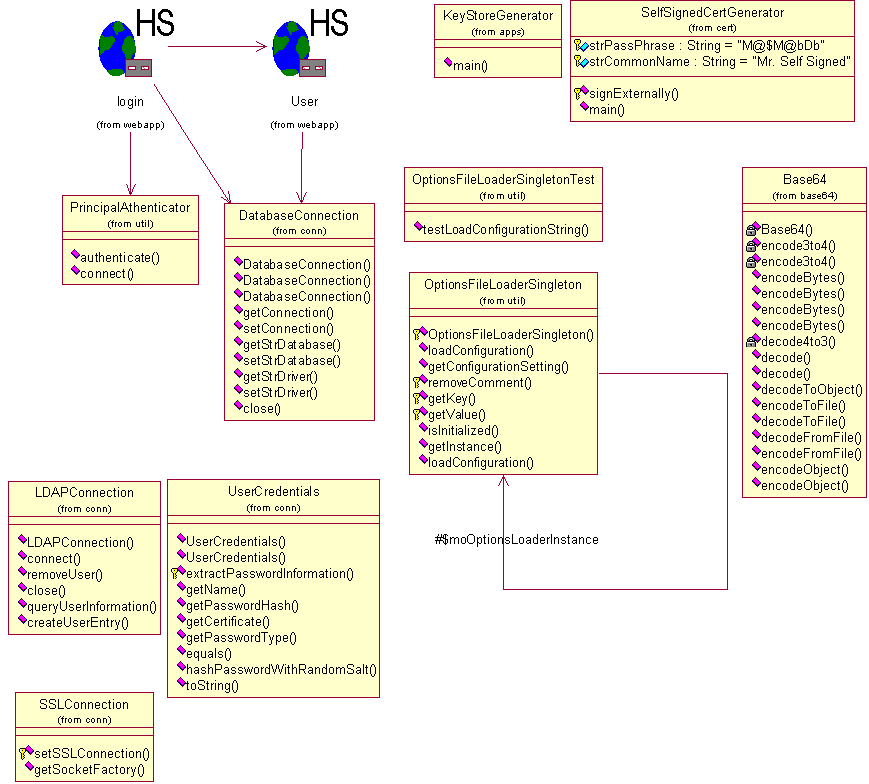}
      }
    }
    \caption{\label{fig:mainClassDiag} SFS main class diagram}
\end{center}
\end{figure}

Please consult Figure \ref{fig:mainClassDiag} for a high-level view
of the class diagram of our application. Please note that the
servlets \texttt{login} and \texttt{User} also have a fair amount of
business logic integrated in them.
This situation is due to the evolutionary nature of the development
method used in this project which, combined with tight deadlines,
did not allow for a proper refactoring of the classes.

We can also take note the presence of test classes in our class
diagram, which are JUnit test cases that allowed to perform some
unit testing. The smallness of the class diagram is mostly explained
by the fact that most of the functionality was implemented in COTS
programs that needed only some configuration.

\section{Class Diagrams}

The following diagrams show some of the important user interfaces of
the SFS software system.

\begin{figure}[htbp]
\begin{center}
  \fbox{
      \scalebox{0.41}{\includegraphics{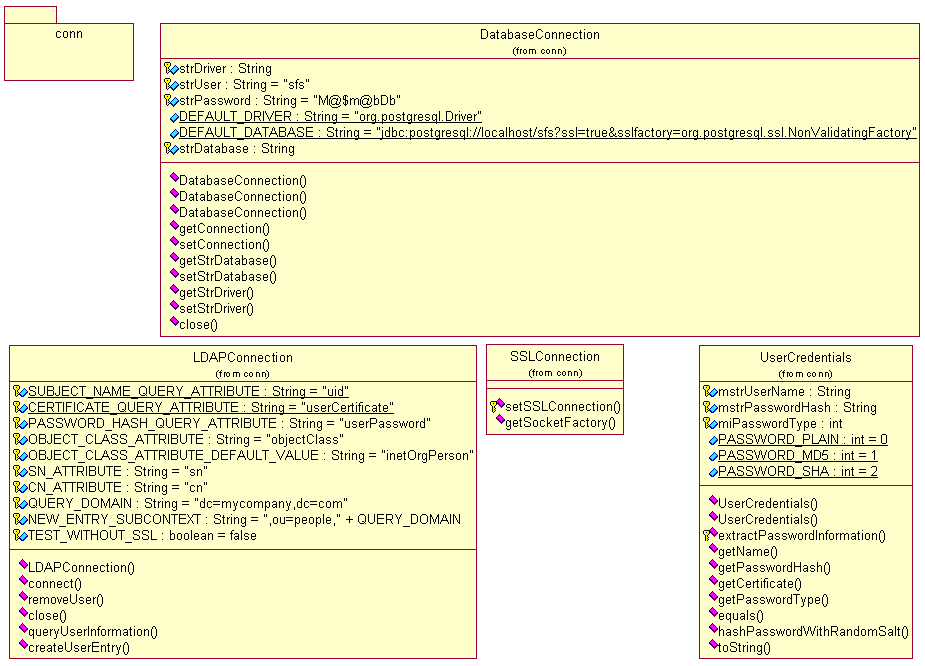}
      }
    }
    \caption{\label{fig:connpackageDiag} Conn Package Diagram}
\end{center}
\end{figure}

\begin{figure}[htbp]
\begin{center}
  \fbox{
      \scalebox{0.41}{\includegraphics{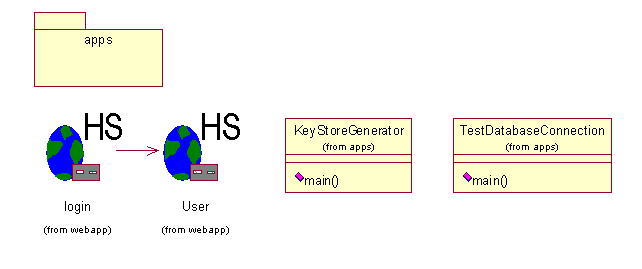}
      }
    }
    \caption{\label{fig:appspackageDiag} Application Package Diagram}
\end{center}
\end{figure}

\begin{figure}[htbp]
\begin{center}
  \fbox{
      \scalebox{0.5}{\includegraphics{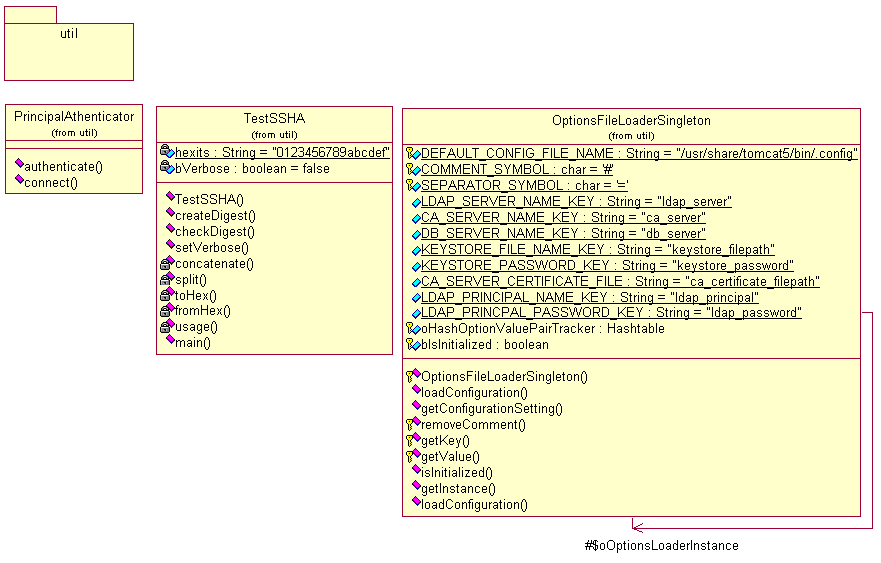}
      }
    }
    \caption{\label{fig:utilpackageDiag} The \texttt{util} Package Diagram}
\end{center}
\end{figure}

\begin{figure}[htbp]
\begin{center}
  \fbox{
      \scalebox{0.41}{\includegraphics{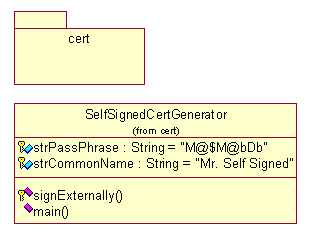}
      }
    }
    \caption{\label{fig:certpackageDiag} The \texttt{cert} Package Diagram}
\end{center}
\end{figure}

\section{User Interface} 
\begin{figure}[htbp]
\begin{center}
  \fbox{
      \scalebox{0.34}{  \includegraphics{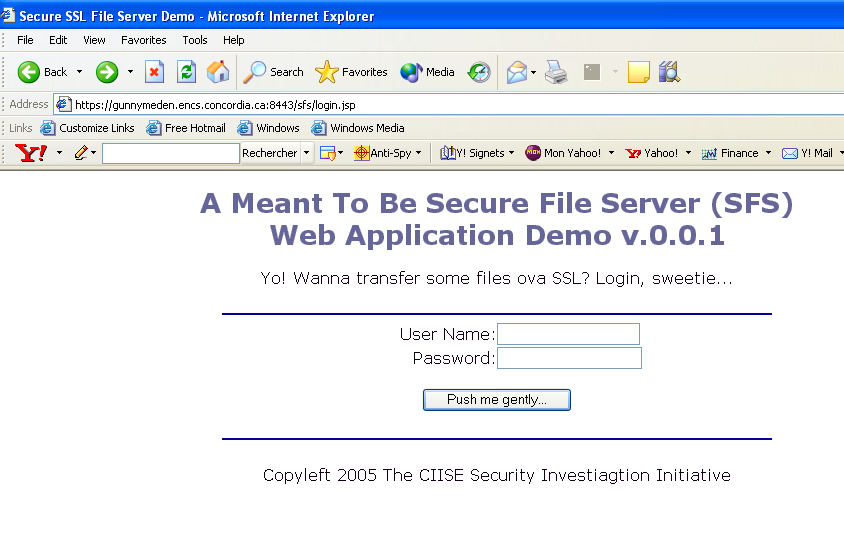}
      }
    }
    \caption{\label{snap1} User interface Log on}
\end{center}
\end{figure}

The following diagrams show some of the
important user interfaces of the SFS software system.
On figure \ref{snap1}, we see the interface allowing clients to log in.\\

\begin{figure}[htbp]
\begin{center}
  \fbox{
      \scalebox{0.47}{  \includegraphics{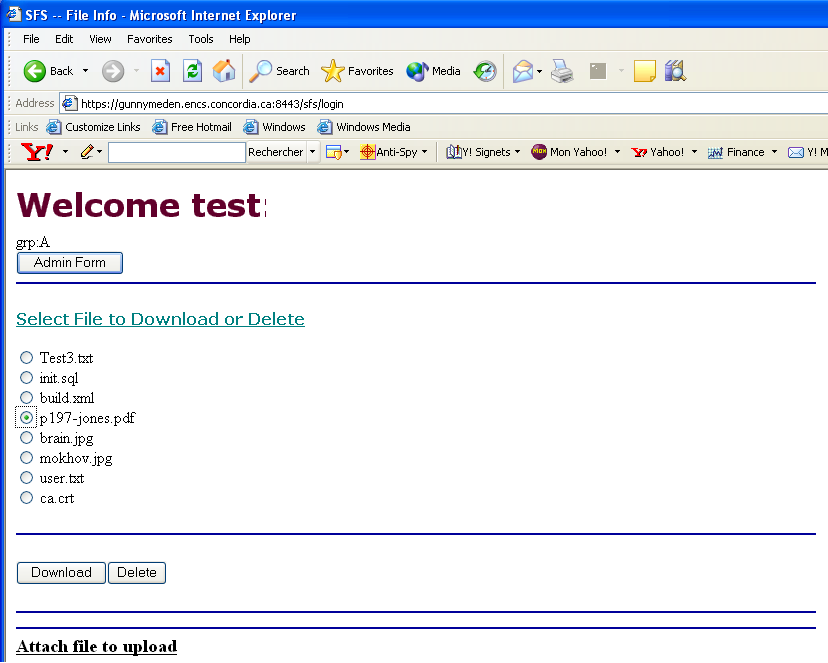}
      }
    }
    \caption{\label{snap2} User operations displayed}
\end{center}
\end{figure}

Once the client mutual authentication is achieved and user allowed
to use the system he will get the following screen (\ref{snap2}).\\

When the client chooses the file to download, he will be prompted to
open the file or give the path he want to save the file in.\\

\begin{figure}[htbp]
\begin{center}
  \fbox{
      \scalebox{0.45}{  \includegraphics{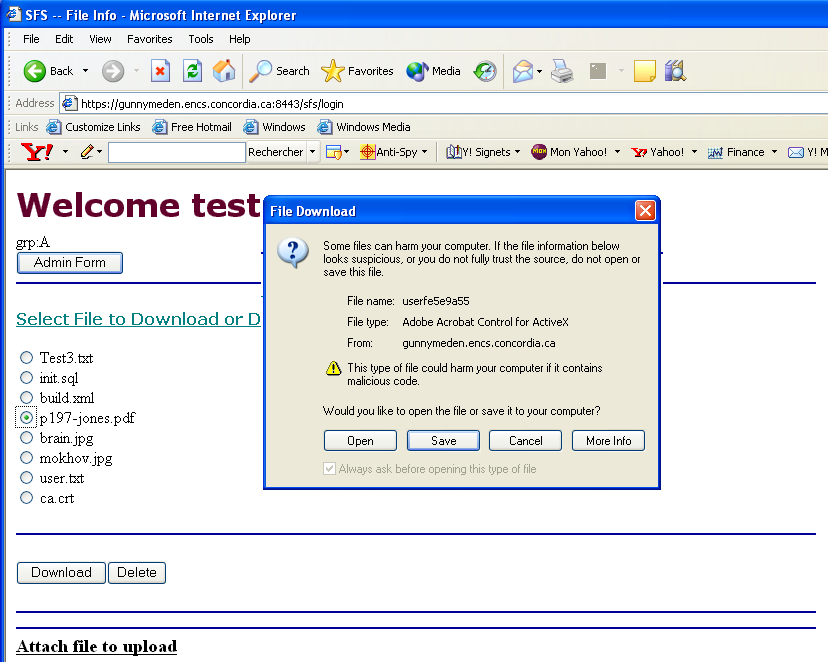}
      }
    }
    \caption{\label{snap3} User operations displayed}
\end{center}
\end{figure}

\begin{figure}[htbp]
\begin{center}
  \fbox{
      \scalebox{0.45}{  \includegraphics{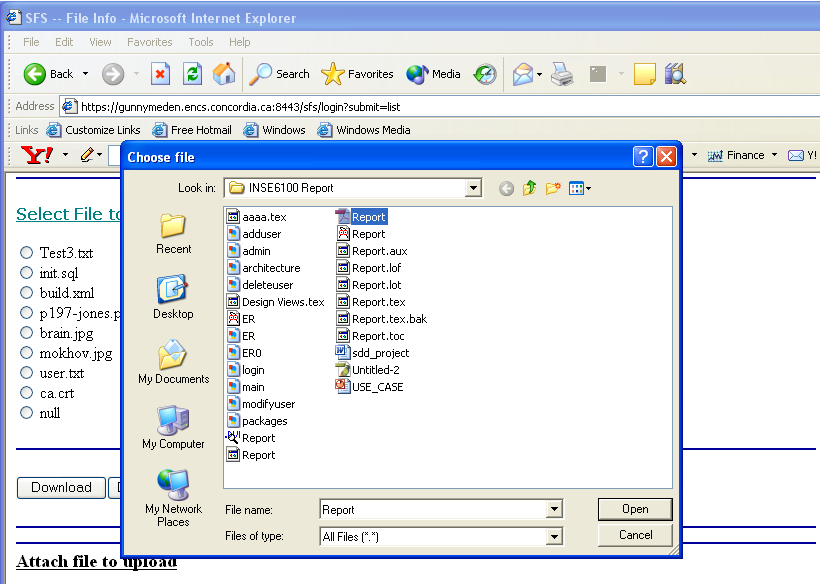}
      }
    }
    \caption{\label{snap4} User operations displayed}
\end{center}
\end{figure}

Figure \ref{snap4} shows the upload operation, so the user will be prompted to
select the file he want to upload.\\

\begin{figure}[htbp]
\begin{center}
  \fbox{
      \scalebox{0.45}{  \includegraphics{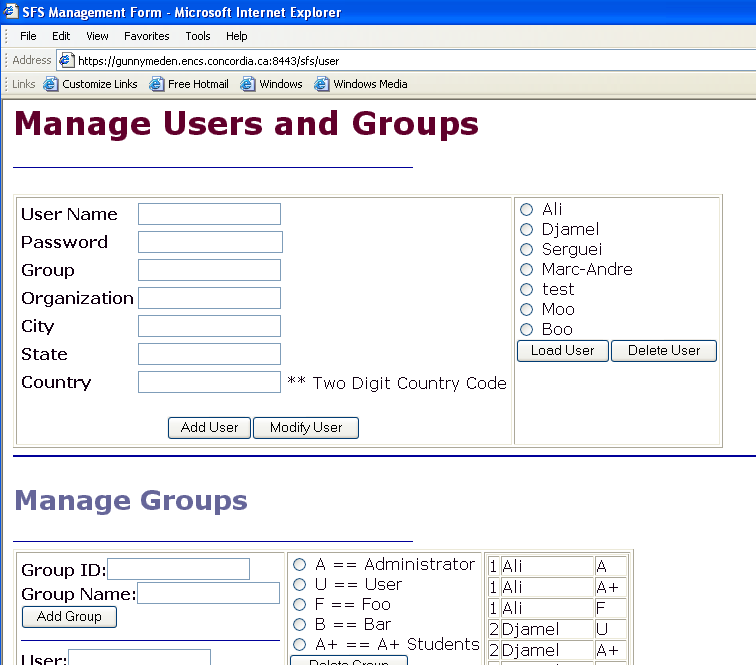}
      }
    }
    \caption{\label{snap5} User operations displayed}
\end{center}
\end{figure}

Figure \ref{snap5} shows the administrator capabilities: adding
users, remove users, adding groups, setting rights, etc.

The snapshot of the SFS provides the login interface via which the
services of the SFS system can not be utilized unless the user is
already logged in.

\section{Class Diagrams}

Figure \ref{fig:mainClassDiag} includes most of the classes already
present. We will describe a few classes in detail here. All the
details regarding the classes are available in the Javadoc.

\subsection{LDAPConnection}
This class provides an abstraction of an LDAP JNDI context, as well
as pre-made queries for obtaining user credentials, deleting an user
and adding an user in the database.

It depends on SSLConnectionFactory for ensuring that our SSL
Connections are set with the proper keys. It also depends on
UserCredentials, since this is the data type it returns on a query.

\subsection{UserCredentials}
This class encapsulates a user name, a password, and a certificate.
It integrates the hashing of plaintext passwords, as well as a
matching comparison between two sets of credentials.
It depends on TestSSHA for generating and validating the salted SHA1 hash.

\subsection{OptionsFileLoaderSingleton}
This class is a singleton, meaning that only up to one instance can
exist at any time. It loads and parses a configuration file. It also
includes many default keys of the configuration file as constant
strings.

\subsection{DatabaseConnection}

This class provides an abstraction for an SSL-enabled database connection.
Contrary to LDAPConnection, it does not provide high-level methods
in it, leaving to the calling code the responsibility to formulate
proper SQL queries.

\section{Data Storage Format}
In this section, we provide a description for the database handling
the security aspect of the system. It consists of the following
entity relation model.

\subsection{Entity Relationship Diagram}

\begin{figure}[htbp]
  \begin{center}
    \fbox{
      \scalebox{0.48}{  \includegraphics{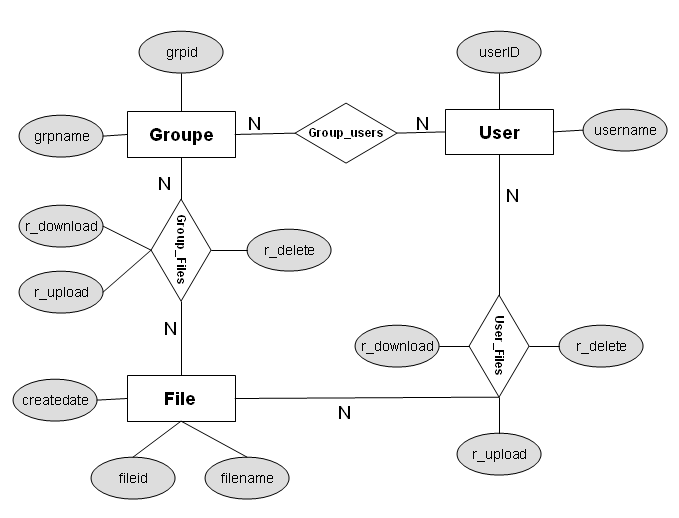}
      }
    }
    \caption{\label{fig:erdiagram} Entity Relationship Diagram}
  \end{center}
\end{figure}

The Groupe entity contains the list of groups a user may belong to. The
User entity contains the list of users having right to use the
system. The File entity contains information about different files a
user can upload, download, delete and view. A user may be an
Administrator or normal user. The other relationships (group\_user,
group\_files) are defined between entities Groupe and User and
Groupe and File to host different information related to both of them
respectively. Hereafter, we provide in Figure \ref{fig:erdiagram}
the Entity Relationship Model of the Security Database.

\section{Options File}

A \texttt{.config} file is expected in the execution root in order
to read the configuration options.

Lines can be comments (\#), blank, or containing a
\texttt{key=value} pair.

The expected configuration options are:

\begin{itemize}

    \item \emph{ca\_server:} host name of the CA server. This option is reserved for future use.
    \item \emph{db\_server:} host name of the database server.
    \item \emph{keystore\_filepath:} relative or absolute path for the web server's keys
    \item \emph{keystore\_password:} password for the previously specified keystore
    \item \emph{ca\_certificate\_filepath:} path to the CA certificate's keystore
    \item \emph{ca\_certificate\_password:} password for the CA certificate
    \item \emph{ldap\_password:} administrator password for the LDAP server
    \item \emph{ldap\_principal:} administrator user name for the LDAP server
    \item \emph{ldap\_server:} host name of the LDAP server

\end{itemize}

\section{Directory Configuration}

The LDAP directory is to be structured in an abritarily manner, as
the  \texttt{DN} is not used in queries. However, the \texttt{UID}
parameter is used for querying based on the user name. The user
information is of type \texttt{inetOrgPerson}, with the fields
\texttt{uid}, \texttt{userPassword} and \texttt{userCertificate}
for, respectively, the user name, its password (hashed with salted
SHA) and its certificate.

\section{External System Interfaces}

The only externally reachable interface to the SFS system is the
login page of the web application. This page should be located at
\texttt{/sfs/login.jsp}. It is also symlinked to it via \texttt{index.jsp}.

\subsection{External Systems and Databases}

The SFS system is not designed for interacting with any other
systems than those described as part of our architecture.

\section{User Scenarios}

\subsection{Typical Scenario}

\begin{enumerate}
\item
User connects to remote server using a web browser on a secure connection
\item
User is prompted with a logon screen, and provides a username and
password
\item
After validation, the user is logged as a normal user and is shown a
list of files to which he or she has access to, as well as their rights
\item
The user clicks to download a file and, if the system validates its
rights, the download begins
\end{enumerate}

\subsection{Variant scenario}
4. The user chooses to upload a file and the upload begins. The file
will be modifyable by the user and according to the default new file ACL.

\subsection{Variant scenario}

4. The user chooses to delete a file to which it has delete rights and
the systems perform the deletion

\section{Administrator Scenarios}
\subsection{Typical Scenario}
\begin{enumerate}
\item
User connects to remote server using a web browser on a secure connection
\item
User is prompted with a logon screen, and provides a username and
password
\item
After validation, the user is logged as an administrator and is shown
a menu of options, and chooses to view a list of files in the system
\item
The user clicks on the user edit button and changes the access
control list of the object.
\end{enumerate}

\subsection{Variant Scenario}
\begin{enumerate}
\item
User connects to the certificate administration service and generate
a certificate for a given subject through a secure connection
\item
User connects to remote server on a secure connection
\item
User is prompted with a logon screen, and provices username and password
\item
After validation, the user is logged as an administrator and is shown
the menu
\item
The user clicks on the certificate edit button and is shown a screen
of certificate maintenance
\item
The user uploads the certificate and the system binds it with the
certificate's defined principal (or creates the user if none exists already)
\end{enumerate}

\subsection{Variant Scenario}
\begin{enumerate}
\item
User connects to remote server using a web browser on a secure connection
\item
User is prompted with a logon screen, and provides a username and
password
\item
After validation, the user is logged as an administrator and is shown
a menu of options, and chooses to edit the certificates
\item
The user chooses to remove the certificate from a given principal
\item
The user connects to the certificate administration service and
issues a certificate revocation.
\end{enumerate}


	\addcontentsline{toc}{chapter}{Bibliography}


\nocite{frequently-used-ssl-commands}

\nocite{tomcat}

\nocite{soen-uml-and-patterns-2006}

\nocite{eclipse}

\nocite{jexplorer}

\nocite{junit}

\nocite{oreilly-servlet-cos}

\nocite{jsp}

\nocite{debbabi-inse6120-200}


\nocite{openssl}

\nocite{postgres}

\nocite{servlets}

\nocite{suranjan03}

\nocite{openldap}

\bibliography{common/report}
\bibliographystyle{alpha}



	\printindex
\end{document}